\documentclass[preprint,12pt]{elsarticle}

\usepackage{amssymb}

\usepackage{amsmath}

\usepackage{hyperref}
\usepackage{color}
\usepackage{subcaption}

 \usepackage{lineno}

\journal{NIM A}

\begin{document}

\begin{frontmatter}

\title{Real time synchronisation of a free-running atomic clock time base with UTC using GNSS signals for application in experimental physics}

\author[a]{Claire Dalmazzone\footnote{Corresponding author: dalmazzone@apc.in2p3.fr}\footnote{Now at Laboratoire AstroParticules et Cosmologie, Université Paris Cité, 4 rue Elsa Morante, 75013 Paris}}
\author[a]{Mathieu Guigue}
\author[a]{Boris Popov}
\author[a]{Stefano Russo}
\author[a]{Vincent Voisin}

\affiliation[a]{organization={Laboratoire de Physique Nucl\'eaire et des Hautes Energies (LPNHE), Sorbonne Universit\'e, CNRS/IN2P3}, addressline={4 place Jussieu}, city={Paris}, postcode={75005}, country={France}}

\begin{abstract}

We present the results obtained by applying, in real-time, a correction method to precisely synchronise a time base generated from a free-running atomic clock with the Coordinated Universal Time (UTC). The method uses the Global Navigation Satellite System (GNSS) signals to have regular time comparisons between the atomic clock generated time base and the GPS Time, perform linear fits of the measurements and extrapolate a correction to apply to the free-running signal. In this work, we apply for the first time this method in real-time. Two atomic clocks were tested, a low-cost Rubidium clock and a more expensive magnetic Caesium clock. We demonstrate that we can obtain a residual difference between the clock time base and the French official realization of UTC (UTC(OP)) in the range of $\pm 15$ ns with no apparent residual drift.
\end{abstract}


\end{frontmatter}


\section{Introduction}

Precise synchronisation of clocks is a growing need in the field of experimental particle physics, especially in long baseline accelerator neutrino experiments where a beam of neutrinos is produced at one site and detected at another site several hundreds of kilometres away, as well as in multi-messenger astrophysics experiments that require a synchronisation to Coordinated Universal Time (UTC). One example is the future Hyper-Kamiokande experiment in Japan~\cite{HK} in which context this work is being performed. This neutrino observatory will have both a long baseline program and a multi-messenger program, performed over a planned time span of at least $20$ years. The local time base will be generated using an atomic clock and synchronised to UTC at the level of $100$~ns using Global Navigation Satellite System (GNSS) signals. 

A common solution for synchronisation to UTC consists in disciplining the frequency of the master clock of the experiment to the signals of a GNSS receiver. In this work, we test a solution where the time base is generated by a free-running atomic clock and the time stamps of the recorded events are corrected using the GNSS signals. This method described in \cite{HKProceedings} and \cite{DALMAZZONE} allows more freedom and control over the setup. For instance, it is more robust against temporary low GNSS exposure or failures of the receiver.

The results presented here are part of studies to prove the feasibility and the efficiency of the developed correction method. A first step towards the validation of this approach was achieved in~\cite{DALMAZZONE} where we demonstrated
the efficiency of the correction in post-processing (applied after the full data-taking
period of $\sim30$ days): a stabilisation of a Rubidium clock drift was achieved at the level of about 5 ns.
The results presented here aim at demonstrating the feasibility and the efficiency of the method in real-time. Indeed, a real-time correction of the time base is needed for instance for the long-baseline program of Hyper-Kamiokande in order to open a data-taking window corresponding to the arrival of the neutrino spill emitted from the J-PARC accelerator $295$~km away. This $5~\mu$s spill is composed of bunches separated by around $600$~ns so a $100$~ns synchronisation is required to see this spill structure. In this paper, the method is applied as an online correction to the signals coming from either a Rubidium clock or a magnetic Caesium one with different intrinsic performances.

\section{Material and Methods}

\subsection{Experimental setup}
\label{sec:setup}

\begin{figure}
    \centering
        \includegraphics[width=\textwidth]{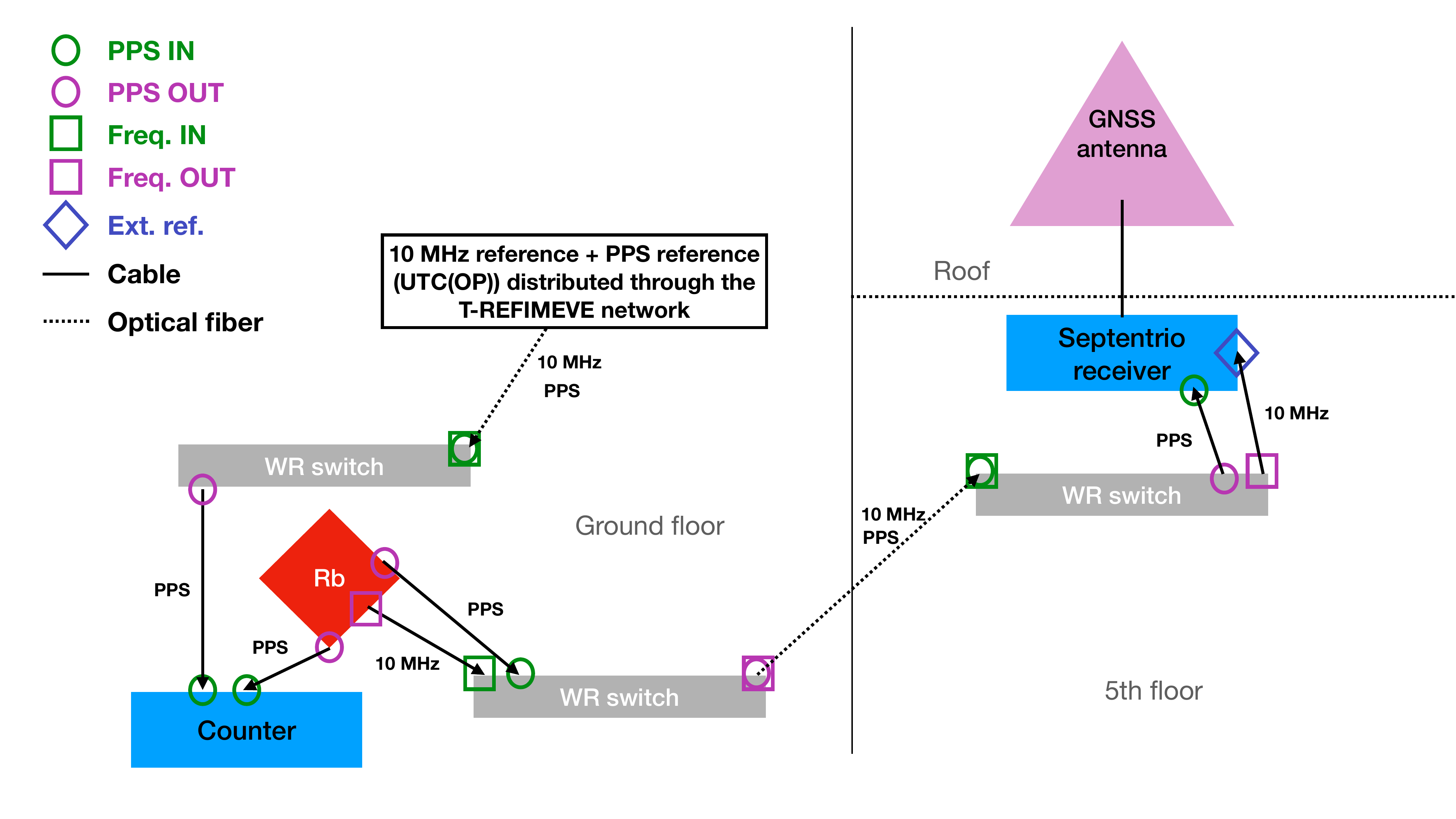}
    \caption{Experimental setup used in this work. A Caesium clock was used instead of the Rubidium clock for the second data taking campaign (red square in the scheme). Figure taken from \cite{DALMAZZONE}.}
    \label{fig:setup}
\end{figure}
 
The same experimental setup as in \cite{DALMAZZONE} was used for this work and is shown in Figure~\ref{fig:setup} for completeness. The atomic clocks are located at the ground floor of our laboratory together with a frequency counter (\href{https://www.keysight.com/us/en/assets/7018-02642/data-sheets/5990-6283.pdf}{53220A} model from Keysight Technologies). In this room, as part of the T-REFIMEVE network~\cite{Refimeve,Refimeve2}, we also have access to the realisation of the UTC from the Observatoire de Paris UTC(OP)~\cite{UTC_OP} and its reference signals, $10$~MHz and Pulse Per Second (PPS), via a White Rabbit (WR) switch~\cite{WR}. We can then measure in real time the difference between the PPS signals of an atomic clock and the UTC(OP). The atomic clock PPS and $10$ MHz signals are also sent to the fifth floor of the laboratory with the WR protocol where we have installed a GNSS antenna (on the roof) and a receiver. Using the atomic clock signals as external reference, the receiver provides a measurement of the time difference between the atomic clock and the GNSS time bases. All the material used was already described in detail in~\cite{DALMAZZONE} except for the Caesium clock, that was acquired later and is presented below.

\subsection{Caesium clock}

\begin{figure}
    \centering
        \includegraphics[width=0.9\textwidth]{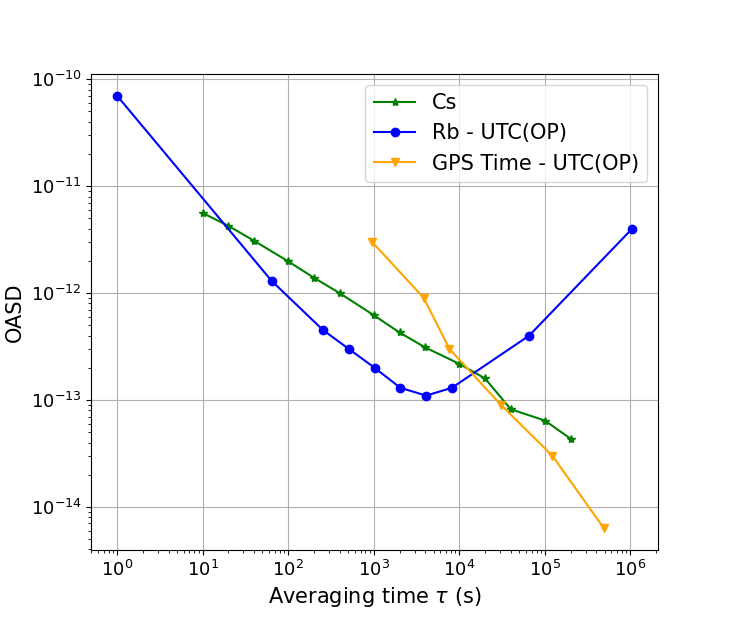}
    \caption{Overlapping Allan standard deviation (OASD) of the Caesium clock frequency signal (green) as measured by the manufacturer, the Rubidium clock PPS signal (blue) and the GPS signals (orange) measured at LPNHE against UTC(OP) using the counter and the GNSS receiver respectively.}
    \label{fig:CsOASD}
\end{figure}

While previous work~\cite{DALMAZZONE} only used a FS725 Rubidium Frequency Standard atomic clock (sold by \href{https://www.thinksrs.com/products/fs725.html}{Stanford Research Systems} with the PRS10 model of Rubidium oscillator),  the correction method has been tested using a Magnetic Caesium atomic clock: the OSA3235B model from the Oscilloquartz company. It provides several frequency signal outputs, including two $10$~MHz signals, and two PPS output signals. Its frequency stability for various averaging times as measured by the factory acceptance tests is shown in Figure~\ref{fig:CsOASD} (green curve).  As a comparison, we also show in blue the stability of the Rubidium clock PPS signal with respect to the UTC(OP) measured in our laboratory and in orange the stability of the GPS signals measured with our GNSS receiver against UTC(OP).  
The Caesium clock OASD keeps decreasing with increasing averaging times up to at least $10^5$~s showing that there is no apparent frequency drift, contrary to the Rubidium clock whose OASD starts increasing for averaging times higher than $4\times10^3$~s because of random walk and linear drift of the frequency.

\subsection{Correction method}

 The correction method we developed was also extensively described in Section 2.2 of Ref.~\cite{DALMAZZONE}. In summary, we use the measurements of time difference between the clock and the GNSS time bases provided every $16$ minutes by the GNSS receiver in CGGTTS format~\cite{CGGTTS} to extrapolate (in online mode) the drift of the clock time with respect to UTC at all times. More precisely, we perform a linear fit of the last $N$ receiver measurements and use the result to extrapolate the near future drift of the atomic clock with respect to UTC. The correction is thus updated every time we receive a new measurement from the receiver. The number $N$ of last receiver measurements to use is optimised depending on the stability of the clock. In particular, for a Rubidium clock whose frequency follows a random walk, the fit window should not be significantly bigger than the characteristic time scale of this random walk and it was determined that the optimal $N_\text{Rb}$ is in the range of $10$ to $30$~\cite{DALMAZZONE}. 
 For the Caesium clock, which does not exhibit any sign of frequency drift, the choice of  $N_\text{Cs}=100$ was motivated by the comparison of the stability of the Caesium clock and the GNSS receiver PPS signal using Figure \ref{fig:CsOASD}. The latter becomes more stable in the range $N=10$ to $N=100$. 

\subsection{Real-time implementation using MIDAS}

 \begin{figure}
    \centering
        \includegraphics[width=\textwidth]{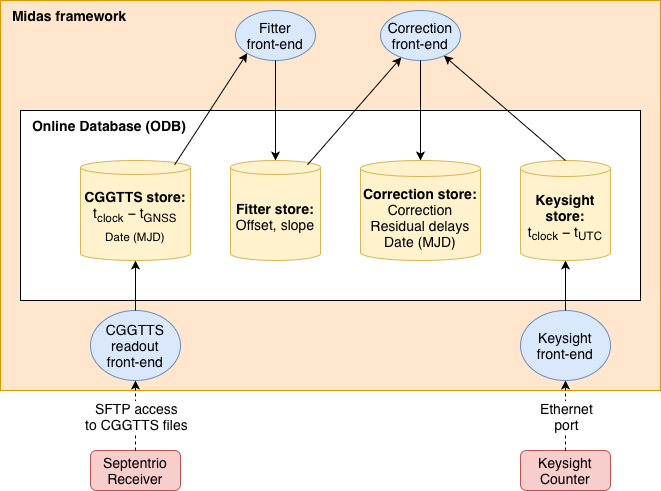}
    \caption{Schematic representation of the MIDAS environment developed for the online monitoring and correction test. It includes four MIDAS front-ends (light-blue ovals) used to control the counter measurements, to access the results of the GNSS receiver's measurements, to update the linear fits of the receiver's measurement and finally to correct the counter's measurements. Each front-end reads data from and writes results to the MIDAS Online DataBase (ODB) in dedicated "stores" (yellow buckets). The solid arrows represent the data flows within the MIDAS environment and the dashed arrows represent the data flow between the MIDAS environment and external instruments.}
    \label{fig:midas}
\end{figure}

 To test the efficiency of the correction in real time, we use the experimental setup described in Section \ref{sec:setup} and the MIDAS data-acquisition software \cite{Midas,Midas2}. 
 The MIDAS ecosystem manages so-called ``front-ends'' controlling and collecting data produced by the GNSS receiver and the frequency counter.
  The dataflow is represented in Figure~\ref{fig:midas}, in particular the recording of the data in the common ``Online Database'' (ODB). The tests were conducted on a Dell Precision M4700 laptop equipped with an Intel Core i5‑3380M processor (2.90 GHz) and an Intel 82579LM Gigabit Network Connection (1 Gbit/s) network interface.

 Compared to the previous work \cite{DALMAZZONE}, data produced by the GNSS receiver in CGGTTS format are read via Secure File Transfer Protocol (SFTP) connection to the receiver and stored in the ODB.
Then the drift of the clock is calculated in real-time using the GNSS receiver's measurements. For this, a dedicated python MIDAS front-end is used to update and save the results of the linear fit of the last $N$ receiver measurements every time a new entry for the GNSS receiver is detected in the database. More precisely, the script reads the new entry, stores the new measurement and its date (to the minutes precision) in a python deque (double ended queue) and rejects the oldest measurement from this deque. This deque thus has a fixed length corresponding to the number of last measurements we want for the linear fits. Then these measurements are fitted to extract a slope $s$ (in seconds per day) and an offset $o$ (in seconds) that best characterise the time evolution of the last GNSS receiver's measurements. These parameters are then stored in the database.

The counter used to compare the clock 1 PPS and the UTC(OP) signals is operated via IEEE 488.2 (SCPI) commands over a socket connection and a MIDAS front-end. The measurements are also read through this protocol  and stored in the ODB. 
 Similarly, every time a new entry $t_\text{clock}-t_\text{UTC}$ is detected for the counter in the database, a ``Correction'' MIDAS front-end reads the current values of the slope $s$ and offset $o$ and computes the correction to be applied to this last counter's measurement $\Delta t = s\cdot D+o$ where $D$ is the approximate date of the counter's measurement. The correction $\Delta t$, the corrected counter's measurement $t_\text{clock}-t_\text{UTC}-\Delta t$ and the date $D$ are stored in the database. The correction is thus applied in quasi-real time with a short delay due to the time required to read from the database and calculate the correction. In the setup used here, this delay is of the order of $1$~s, but it was optimised since then. This limitation is further discussed in Section~\ref{sec:discussion_speed}. In summary, it only limits the amount of data that can be treated per unit of time without impacting the accuracy of the correction.

\section{Results}

\subsection{Performance of the correction}

 \begin{figure}[!ht]
    \centering
        \includegraphics[width=\textwidth]{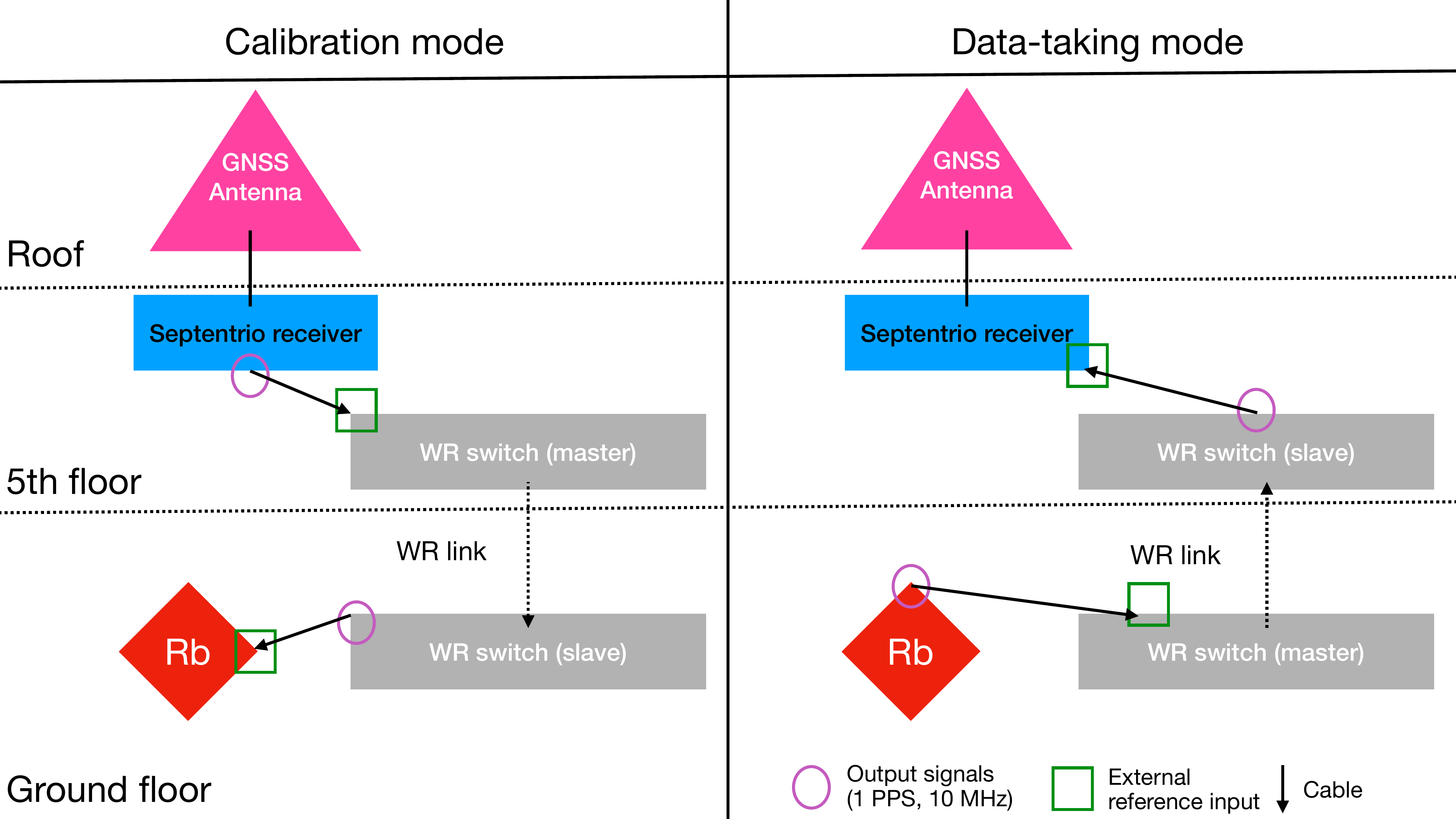}
    \caption{Schematic representation of the Rubidium clock - GNSS receiver part of the setup in calibration (left) and data-taking (right) modes. In calibration mode, the GNSS receiver uses the GNSS signals to generate a 1 PPS and $10$~MHz signals that are used as reference by the fifth floor WR switch that is in ``master mode'' in this setup. The ground floor WR switch is thus the slave switch and its output $10$~MHz is used to discipline the Rubidium clock. In data-taking mode, the master switch is at the ground floor and it uses the undisciplined Rubidium clock signals as reference. The output signals of the slave switch are used as reference for the GNSS receiver to generate a 1 PPS signal to compare to the GNSS signals.}
    \label{fig:calib}
\end{figure}

First, we used the setup with the Rubidium clock for around $14$ days. Before the data-taking, the Rubidium PPS signal was aligned with the GNSS receiver PPS. To do so, the WR link from the ground floor to the fifth floor was inverted. The differences between the ``calibration'' and the ``data-taking'' modes of the setup are illustrated in Figure~\ref{fig:calib}. The calibration mode allows to use the White Rabbit link from the WR system on the fifth floor (here called master switch) to transmit time to the WR system on the ground floor (here the slave switch). The 10~MHz output from the slave WR system was then used to discipline the Rb clock for a few days.  
Then the WR link was inverted to revert to the original setup (data-taking mode) where the GNSS receiver can be used to measure the time difference between the Rubidium clock and the GNSS 1 PPS. Also, the counter was used to measure the time difference between the clock (in free-running mode) and the UTC(OP) PPS signals every $10$ seconds. The default setting of the counter allows one measurement per second, however the length of time required to compute the correction means that not all events can be corrected and stored at this counter frequency. For this reason we took only one measurement every $10$ seconds. For the test with the Caesium clock which lasted for 80 days, it was decided to take one counter measurement per second despite this limitation. This will be further discussed in Section~\ref{sec:discussion_speed}.

\begin{figure}
    \centering
    \begin{subfigure}{0.48\textwidth}
        \includegraphics[width=\textwidth]{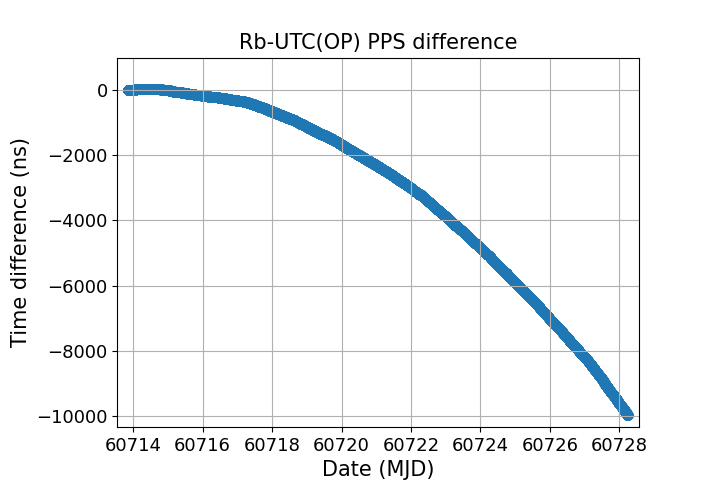}
        \caption{Rubidium clock: $\sim14$ days}
    \end{subfigure}
    \begin{subfigure}{0.48\textwidth}
        \includegraphics[width=\textwidth]{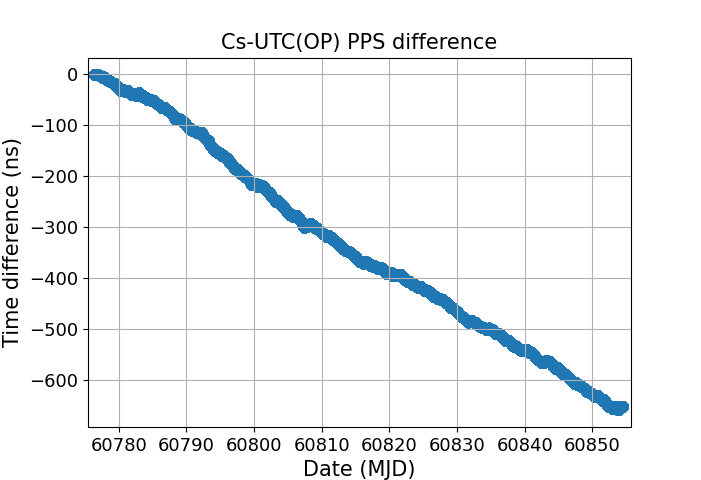}
        \caption{Caesium clock: $\sim80$ days}
    \end{subfigure}
    \caption{Time differences between the PPS of the atomic clock (Rb on the left and Cs on the right) and the UTC(OP) measured by the counter. For the Rb clock (left) the counter took one measurement every ten seconds whereas for the Cs clock it took one measurement per second.}
    \label{fig:free_running}
\end{figure}

The counter measurements for both clocks are shown in Figure~\ref{fig:free_running}: they confirm the need for a correction of a time base generated from a free-running atomic clock if a $100$~ns synchronisation to UTC is required. Without such a correction, a $100$~ns drift was obtained in around $1$ day for the Rubidium clock and $14$ days for the Caesium clock. Also, as expected from the frequency stability of the two systems, the long term drift of the Rubidium clock seems to be quadratic in time whereas it looks linear for the Caesium clock. The drift of the Caesium clock time with respect to UTC is attributed to the limited precision of its calibration. The Caesium clock's frequency was calibrated with respect to UTC reference by the manufacturer with a $10^{-13}$ relative precision. However small, a difference of frequency with respect to UTC will manifest as a linear drift of the clock's PPS signal with respect to the UTC PPS signal. The measured drift fitted with a linear function demonstrates a precision of calibration of $\sim10^{-13}$ which is within the accuracy advertised by the manufacturer. For the Rubidium clock, the long term drift of the PPS signal is dominated by the frequency drift, hence the quadratic drift of time. 

\begin{figure}
    \centering
    \begin{subfigure}{0.48\textwidth}
        \includegraphics[width=\textwidth]{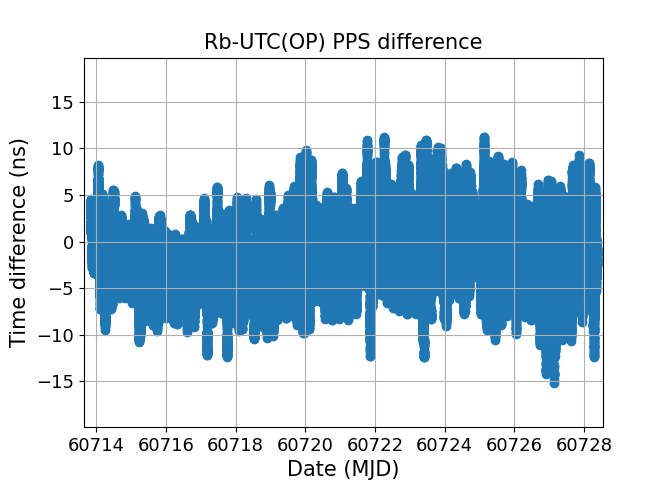}
        \caption{Rubidium clock: $\sim14$ days}
    \end{subfigure}
    \begin{subfigure}{0.48\textwidth}
        \includegraphics[width=\textwidth]{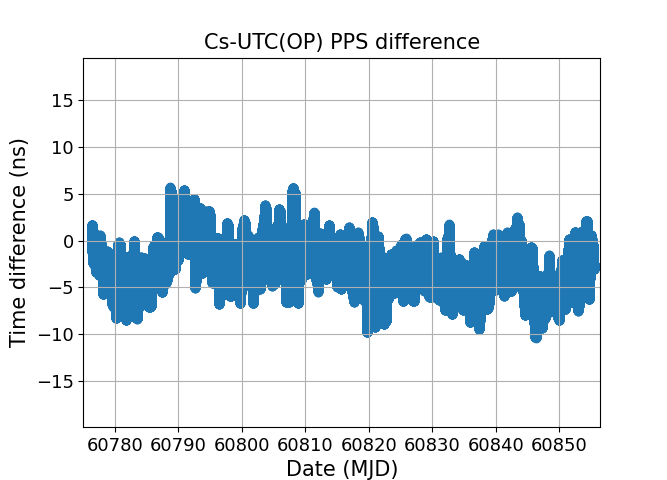}
        \caption{Caesium clock: $\sim80$ days}
    \end{subfigure}
    \caption{Residual time differences between the PPS of the atomic clock (Rb on the left and Cs on the right) and the UTC(OP) after the correction applied in real time on the measurements of the counter. The x-axis corresponds to the date and time (in Modified Julian Day) at which the correction was applied.}
    \label{fig:residuals}
\end{figure}

Figure \ref{fig:residuals} shows the residual difference between the atomic clocks and the UTC(OP) time bases after correction as a function of the date at which the correction was applied. Both distributions show that there is no apparent residual long term drift in the corrected signals and the residuals stay within a range of $\pm15$ ns for both clocks with a standard deviation of $2.8$~ns for the Rubidium clock and $2.4$~ns for the Caesium clock.

\subsection{Stability of corrected signals}
    \label{sec:appendix}

    \begin{figure}[!ht]
        \centering
        \includegraphics[width=\textwidth]{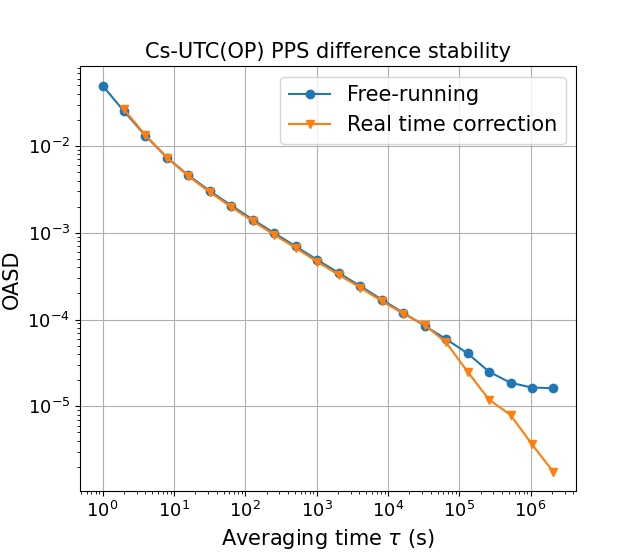}
        \caption{Overlapping Allan standard deviation (OASD) of the free-running Caesium clock PPS signal (blue) as measured by the counter and the signal corrected in real time (orange) with respect to UTC(OP).}
        \label{fig:OASD_afterCorr}
    \end{figure}

    The stability of the Rb-UTC(OP) PPS difference before and after the correction for the Rubidium clock was already extensively described in~\cite{DALMAZZONE}. This section describes only the stability of the Cs-UTC(OP) PPS difference in free-running and corrected modes as shown in Figure~\ref{fig:OASD_afterCorr}. It demonstrates that the correction does not significantly deteriorate the short term stability of the signal and actually improves the long term stability. Indeed, for averaging times above $8\times10^4$~s, the OASD of the corrected signal decreases faster with increasing averaging times than the OASD of the free-running signal because it follows the stability of the GNSS signals that is not limited by flicker noise like the Caesium clock. This switch happens at the time corresponding to the integration time window of the correction: $N_\text{Cs}=100$ receiver's measurements correspond to $\sim 10^5$~s of integration of GNSS signals.  

    While the first OASD value of the free-running signal is at an averaging time of $1$~s, the first OASD value of the corrected signal corresponds to an averaging time slightly above $1$~s because the event rate in the corrected signal was slightly reduced compared to the free running signal. This is due to the problem discussed in Section \ref{sec:discussion_speed} that the running time of the correction is too big compared to the counter's measurement rate. 

\section{Discussion}

\subsection{Tolerance of the GNSS receiver}
\label{sec:discussion_jump}

\begin{figure}[!ht]
        \centering
        \includegraphics[width=\textwidth]{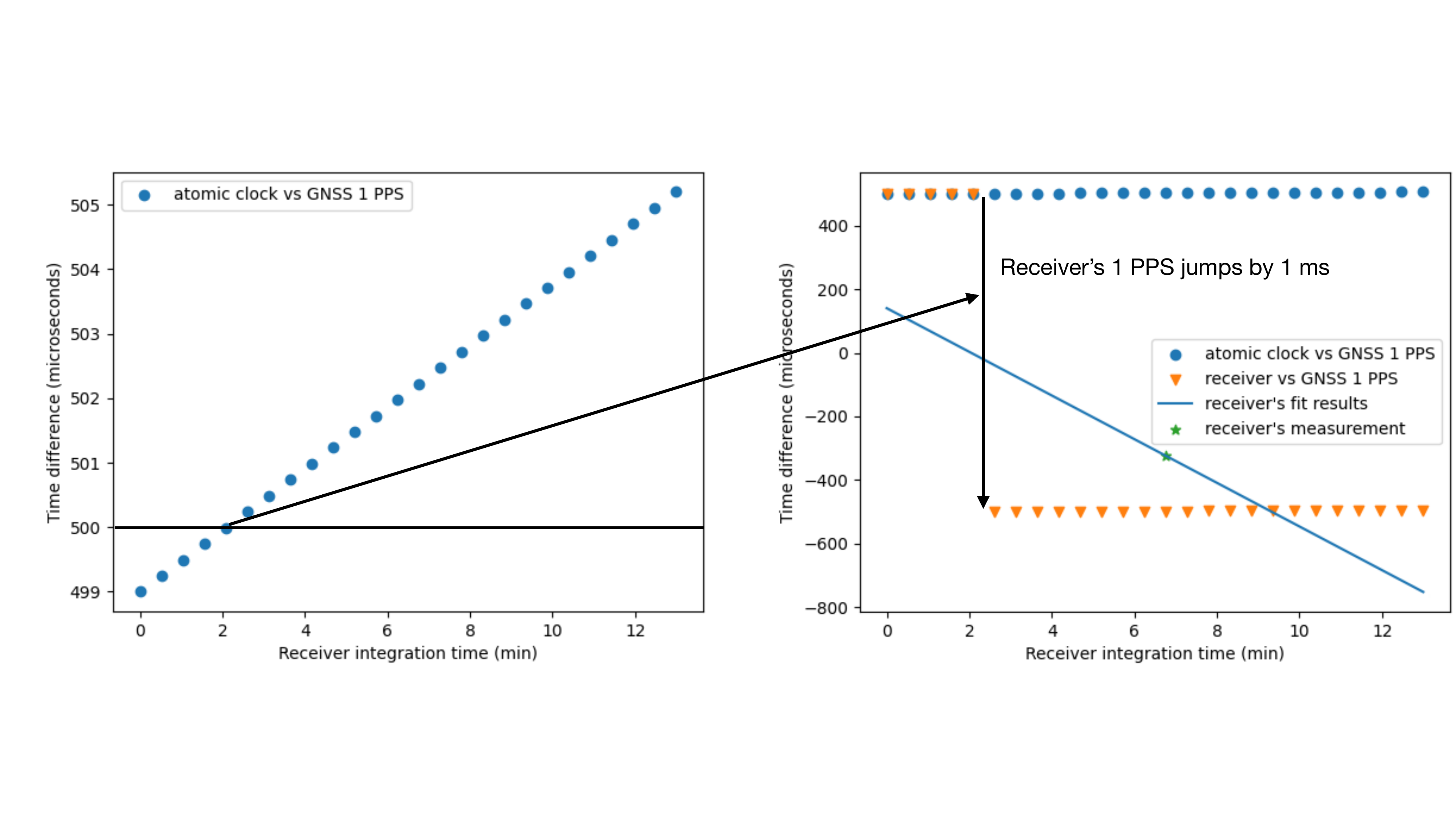}
        \caption{Schematic representation of what happens when the absolute difference between the atomic clock 1 PPS and the GNSS 1 PPS becomes greater than $500~\mu$s. The receiver's 1 PPS signal then jumps by $1$~ms and the measurement of the corresponding integration time window is not reliable. For the following measurements, the result will be reliable but with a $1$~ms offset.}
        \label{fig:receiver_jump}
    \end{figure}

The GNSS receiver can measure the time difference between an external reference (here the atomic clock) and the GNSS Time using the external $10$~MHz signal to generate  its local time base and the external PPS signal to initially align its PPS. To do so, it tracks each satellite in view during around $13$ minutes, performs a linear fit of the measured drift between its own 1 PPS and the satellite's one and returns the result of the fit at the centre of the integration time window of $13$ minutes. However, there exists a certain tolerance for the difference between the two time bases. For our receiver, with the default settings, this tolerance is $\pm 500~\mu$s. If the time difference exceeds this limit, the receiver's 1 PPS will jump by $\pm1$~ms so that the difference between the receiver and the GNSS 1 PPS falls back into the tolerance range. From this moment on, the receiver's 1 PPS signal presents a constant $1$~ms offset compared to the 1 PPS signal from the atomic clock. Also, as illustrated by Figure \ref{fig:receiver_jump}, the measurement provided by the receiver just after this jump is not reliable because the jump will have happened within the corresponding integration time window. 

This jump is purely due to the receiver's properties, not the atomic clock signals, so it should not be propagated to the correction we apply. As a consequence, the script that computes the correction should be able to recognise that there was such a jump and discard the unreliable measurements. In particular when using a Rubidium clock, due to its frequency drift, such jumps are expected to happen quite often during the typical lifetime of a particle physics experiment like Hyper-Kamiokande (at least $20$ years). For instance, using Figure~\ref{fig:free_running}, we estimate that the Rubidium clock PPS signal will have drifted by $1$~ms after a few months. After this first jump, due to the quadratic nature of the PPS signal's drift, subsequent jumps will happen more and more frequently. This renders the Rubidium clock challenging to use in free-running mode for long lasting experiments. A compromise would be to monitor the frequency drift and correct it in real time making use once again of the GNSS signals. Indeed, with the SRS FS725 we are using here, the user can change the frequency of the clock with a simple command. This could also be done by a script via the MIDAS environment so that the Rubidium clock would be used in a semi-free-running mode. 

The Caesium clock shows no sign of frequency drift so that a $1$~ms drift of the time base would take several hundred years. Consequently, if the clock's PPS is initially aligned with GNSS Time, we have no reason to expect a jump of the receiver's PPS within the lifetime of the experiment. A careful monitoring of the value of the time difference measured by the GNSS receiver should be enough to make sure no such jump will happen.  

\subsection{Implementation in a particle physics experiment}
\label{sec:discussion_speed}
The real time correction applied here is actually slightly delayed because of the time required to transfer the data to and from the database and the time to calculate the correction and to apply it. This delay is not a problem per se because there is no need for an instant knowledge of the real timing of the event.  However, in our current setup, the delay introduces dead time in the correction process. As a consequence of the limited speed of the process, when having an event rate of $1$ counter measurement per second, some of these measurements were missed due to the dead-time. While this does not impact the demonstration of the efficiency of the correction, this could cause issues in the implementation for a particle physics experiment where the event rate can be much higher\footnote{More than $50,000$ events are expected in Hyper-Kamiokande in a few tens of seconds in case of a supernova neutrino burst at $10$~kpc from the Earth.}. This correction will ultimately be used to correct the time stamps of real physics events, like a neutrino interaction in case of the Hyper-Kamiokande experiment. One 
possible optimisation is to use a MIDAS tool called manalyzer that provides the possibility for direct analysis of the measurements instead of reading them from the ODB. This drastically reduces the dead-time.

\section{Conclusions}

In this paper we demonstrated the efficiency of the method described in~\cite{DALMAZZONE}, when applied in real time, to correct the time base generated from a free-running atomic clock in order to synchronise it with UTC. The Hyper-Kamiokande experiment requires a timing system with timestamps aligned to UTC at an accuracy of better than $100$~ns, meaning that at all times the absolute time difference between the experiment time base and the UTC should be smaller than $100$~ns. In particular for the long-baseline program of the experiment, this level of synchronisation is needed in real-time. The method consists in using a GNSS receiver and the $10$~MHz signal of the atomic clock as an external reference to compare the atomic clock time base to the GNSS Time. The GNSS Time is a good representation of UTC to a few nanoseconds. The receiver measurements are used to extrapolate the drift of the clock with respect to UTC in the near future. This extrapolation is updated every time a new measurement is available from the receiver. This allows, with careful optimisation, to also synchronise clocks whose frequency signals present random walk behaviour like common Rubidium standards. 

The precision of this method was demonstrated using a measurement of the difference between the atomic clock PPS signal and the UTC(OP) PPS reference signal that we have access to through the T-REFIMEVE network. The correction was applied in real time to this measurement provided by a frequency counter thanks to the MIDAS data acquisition framework. The setup was tested with two atomic clocks: a Rubidium SRS FS725 ($14$ days of data-taking) and a Caesium OSA3235B ($80$ days of data-taking). For both clocks, a synchronisation to UTC better than $15$~ns was obtained for the whole data-taking period with standard deviation of the residuals below $3$~ns. This is well within the usual requirements for long baseline and multi-messenger particle physics experiments. However, we also discussed in Section \ref{sec:discussion_jump} the problems arising from the fast drift of the Rubidium clock and the tolerance of the GNSS receiver. For the purpose of running a particle physics experiment continuously for at least two decades, using the more stable magnetic Caesium clock would present the advantage of not having to monitor for a possible jump in the GNSS receiver PPS signal.

\vspace{6pt} 

\paragraph{\textbf{Fundings}}
This research was funded by IN2P3/CNRS, the French "Agence nationale pour la recherche" under grant number ANR-21-CE31-0008, the "IdEx Sorbonne Université" and the 2019 "Sorbonne Université Émergences: MULTIPLY” grant.\\
The White Rabbit network and the access of associated optical fibers to the Pierre and Marie Curie campus: T-REFIMEVE, FIRST TF and LNE: "Agence Nationale de la Recherche" (ANR-21-ESRE-0029 / ESR/Equipex T-REFIMEVE, ANR-10-LABX48-01 / Labex First-TF);  Laboratoire National d’Essai (LNE), project TORTUE.

\paragraph{\textbf{Acknowledgments}}
We acknowledge the useful corrections to the draft version of this article made by Dr Vladimir Gligorov. We also acknowledge the contribution of colleagues from the LNE-SYRTE laboratory, in particular Michel Abgrall, Baptiste Chupin, Caroline B. Lim, Paul-Éric Pottie and Pierre Ulrich, for their help with the inclusion of our laboratory in the T-REFIMEVE network as well as for the calibration of the White Rabbit switches and GNSS receiver used in this work. We also thank the Hyper-Kamiokande collaboration for their continuous support of our research.


\begin{thebibliography}{99}

\bibitem[Hyper-Kamiokande Proto-Collaboration(2018)]{HK}
K.~Abe et al., \emph{Hyper-Kamiokande Proto-Collaboration,  Hyper-Kamiokande Design Report}, (2018), arXiv:1805.04163.

\bibitem {HKProceedings} 
L.~Mellet, M.~Guigue, B.~Popov, S.~Russo, V.~Voisin, on behalf of the Hyper-Kamiokande Collaboration, 
\emph{Development of a Clock Generation and Time Distribution System for Hyper-Kamiokande}, 
\emph{Phys. Sci. Forum} {\bf 8} (2023) 72, https://doi.org/10.3390/psf2023008072.

\bibitem{DALMAZZONE}
C.~Dalmazzone, M.~Guigue, L.~Mellet, B.~Popov, S.~Russo, V.~Voisin, M.~Abgrall, B.~Chupin, C.~B.~Lim, E.~Pottie, P.~Ulrich, \emph{Precise synchronisation of a free-running Rubidium atomic clock with GPS Time for applications in experimental particle physics}, \emph{Nucl.Instrum.Meth.A} {\bf 1075} (2025) 170358, https://doi.org/10.1016/j.nima.2025.170358.

\bibitem{Refimeve}
E.~Cantin et al., 
\emph{REFIMEVE Fiber Network for Time and Frequency Dissemination and Applications}, 2023 Joint Conference of the European Frequency and Time Forum and IEEE International Frequency Control Symposium (EFTF/IFCS), Toyama, Japan, 2023, pp. 1-4, https://doi.org/10.1109/EFTF/IFCS57587.2023.10272084. 


\bibitem{Refimeve2}
C.~B.~Lim et al., 
\emph{Extension of REFIMEVE with a White Rabbit Network}, 2023 Joint Conference of the European Frequency and Time Forum and IEEE International Frequency Control Symposium (EFTF/IFCS), Toyama, Japan, 2023, pp. 1-4, https://doi.org/10.1109/EFTF/IFCS57587.2023.10272069. 

\bibitem{UTC_OP}
G.~D.~Rovera et al., 
\emph{UTC(OP) based on LNE-SYRTE atomic fountain primary frequency standards}, 
\emph{Metrologia} {\bf 53} (2016) S81.

\bibitem{WR}
J.~Serrano et al., 
\emph{The White Rabbit project} (2013), https://cds.cern.ch/record/1743073.

\bibitem{CGGTTS}
P.~Defraigne, G.~Petit, 
\emph{CGGTTS-Version 2E: an extended standard for GPS Time Transfer}, 
\emph{Metrologia} {\bf 52} (2015), 
IOP Publishing, https://doi.org/10.1088/0026-1394/52/6/G1.

\bibitem{Midas}
S.~Ritt, P.~Amaudruz, K.~Olchanski, \emph{MIDAS (Maximum Integration Data Acquisition System)}, https://bitbucket.org/tmidas/midas.

\bibitem{Midas2}
S.~Ritt and P.~A.~Amaudruz, \emph{New components of the MIDAS data acquisition system}, 1999 IEEE Conference on Real-Time Computer Applications in Nuclear Particle and Plasma Physics. 11th IEEE NPSS Real Time Conference. Conference Record (Cat. No.99EX295), Sante Fe, NM, USA, 1999, pp. 116-118, doi: 10.1109/RTCON.1999.842578.



\end{thebibliography}
\end{document}